\documentstyle[prl,aps,psfig]{revtex}
\newcommand{\be}{\begin{eqnarray}}
\newcommand{\ee}{\end{eqnarray}}

\newcommand{\bi}{\begin{itemize}}
\newcommand{\ei}{\end{itemize}}
\begin{document}
\twocolumn[\hsize\textwidth\columnwidth\hsize
           \csname @twocolumnfalse\endcsname
\title{On a correspondence between classical and
  quantum particle systems}
\author{Klaus Morawetz}
\address{
Max-Planck-Institute for the Physics of Complex Systems, 
Noethnitzer Str. 38, 01187 Dresden, Germany}
\maketitle
\begin{abstract}
An exact correspondence is established between a $N$-body classical
interacting system and a $N-1$-body quantum system with respect to the
partition function. The resulting quantum-potential is a $N-1$-body
one. Inversely the Kelbg potential is reproduced which describes
quantum systems at a quasi-classical level. 
The found correspondence between classical and quantum systems allows 
also to approximate dense classical many
body systems
by lower order quantum perturbation theory replacing Planck's constant
properly by temperature and density dependent expressions. As an
example the dynamical behaviour of an one - component plasma is well
reproduced concerning the formation of 
correlation energy after a disturbance utilising solely the
analytical quantum - Born result for dense degenerated Fermi systems. As
a practical guide the quantum - Bruckner parameter $r_s$ has been replaced by
the classical plasma parameter $\Gamma$ as $r_s\approx0.3 \Gamma^{3/2}$.
\end{abstract}
\vskip2pc]

Several hints in recent literature conjecture that there seem to
exist a correspondence between quantum systems and higher dimensional
classical systems. The authors of \cite{BMM00}  argue that a higher dimensional
classical non-Abelian gauge theory leads to a lower dimensional quantum
field theory in the sense of chaotic quantisation. The correspondence
has been achieved by equating the temperature characterising
chaotization  of the higher dimensional
system with $\hbar$ of the lower dimensional system by
\be
\hbar =a T.
\label{1}
\ee
Recalling imaginary time evolution as a method to calculate
correlated systems in equilibrium such correspondence seems
suggestible. We will find a similar relation as a best fit
of quantum - Born calculations to dense interacting classical systems.

In condensed matter
physics it is a commonly used trick to map a two - dimensional classical
spin system onto a one - dimensional quantum system \cite{Ch99}. This
suggests that there might exist a general relation between classical
and higher dimensional quantum systems. We will show that a
classical many body system can be equally described by a quantum
system with one particle less in the system but with the price of
complicated nonlocal potential. This can be considered analogously to
the Bohm interpretation of quantum mechanics \cite{BH84} where the Schroedinger
equation is rewritten in a Hamilton-Jacobi equation but with a
nonlocal quantum potential.

Another hint towards a
correspondence between classical and quantum systems was found recently
in \cite{MM00} where it was achieved to define a Lyapunov exponent in
quantum mechanics by employing the marginal distribution which is a 
representation of Wigner function in a higher dimensional space. Since
the Lyapunov exponent is essentially a concept borrowed from classical
physics this finding points also in the direction that there exists a
correspondence between quantum systems and higher dimensional
classical systems.

On the opposite side there are systematic derivations of
constructing effective classical potentials such that the many body quantum
system is described by the classical system. An example is the Kelbg 
potential for Coulomb systems \cite{K64,K641,EHK67,KK68}
\be
V^{\rm Kelbg}_{12}(r)={e_1 e_2 \over r}
\left (
1-{\rm e}^{-r^2/l^2}+\sqrt{\pi}{r\over l} {\rm erfc}
\left ({r\over l}\right ) 
\right )
\label{r}
\ee
with $l^2=\hbar^2/2 \mu T$ and $1/\mu=1/m_1+1/m_2$
describing the two-particle quantum Slater sum correctly by a
classical system. Improvements and systematic applications 
can be found in \cite{kker86,KK681,OVE00}.

Here in this paper it should be shown that a
classical 
$N$-particle system can be mapped exactly on a
quantum $N-1$-particle system
with respect to the partition function. Though the resulting effective
$N-1$ body quantum potential is highly complex it can lead to practical
applications for approximating strongly correlated classical
systems. In the thermodynamical limit it means that the dense
classical system can be described alternatively by a quantum system
with properly chosen potential. 

This finding suggests that the quantum calculation in
lowest order perturbation might be suitable to derive good
approximations for the dense classical system.
This is also motivated by an intuitive picture. Assume we have a
dense interacting classical plasma system. Then the correlations will restrict
the possible phase space for travelling of one particle considerably
like in dense Fermi systems at low
temperatures where the Pauli exclusion principle restrict the phase
space for 
scattering. Therefore we might be able to
describe a dense interacting classical system by a perturbative
quantum calculation  when properly
replacing $\hbar$ by density and temperature expressions.  Indeed we
will demonstrate in a one - component plasma system that even the time
evolution and dynamics of a very strongly correlated classical system
can be properly approximated by quantum - Born calculations replacing
the quantum parameters by proper classical ones.

Let us now start to derive the equivalence between classical and
quantum systems by rewriting the classical N-particle partition
function. The configuration integral reads
\be
Q_N(\beta)=\int dx_1...dx_N \prod\limits_{i<j}^N (1+f_{ij})
\label{qn}
\ee
where we used Meyer's graphs $f_{ij}=\exp{(-\beta u_{ij}(x_i-x_j))}-1$
with the interaction potential $u_{ij}(x_i-x_j)$ of the classical particles
 and the inverse temperature $\beta$.
It is now of advantage to consider the modified configuration integral
\be
&&\tilde Q_N(\beta)=Q_N(2 \beta)\nonumber\\
&&= \int dx_1...dx_N dx_1'...dx_N' \delta(x_1-x_1')...\delta(x_{N-1}-x_{N-1}')\nonumber\\&&
\matrix{
\times & (1+f_{12})(1+f_{13})(1+f_{14})...(1+f_{1N})\cr
\times & (1+f_{21'})(1+f_{23})(1+f_{24})...(1+f_{2N})\cr
\times &... ...\cr
\times & (1+f_{N1'})(1+f_{N2'})(1+f_{N3'})...(1+f_{NN-1'})
}
\label{q}
\ee
such that a quadratic schema in $f_{ij'}$ appears.
Now we assume a complete set of $N-1$ particle wave functions $\Psi_{n_{N-1}}$
such that
\be
&&\delta(x_1-x_1')...\delta(x_{N-1}-x_{N-1}')\nonumber\\&&=\sum\limits_{i_1..i_{N-1}}
\Psi^*_{i_1..i_{N-1}}(x_1'...x_{N-1}') \Psi_{i_1..i_{N-1}}(x_1...x_{N-1})
\ee
with some "quantum numbers" $\{i\}$ characterising the state. Further we propose the following
eigenvalue problem defining the wave function
\be
&&\int dx_1 \prod\limits_{j=2}^N (1+f_{1j})
\Psi_{i_1..i_{N-1}}(x_1...x_{N-1})
\nonumber\\&&
\qquad\qquad =V
{\rm e}^{-\varepsilon_{\{i\}}} \Psi_{i_2..i_{N-1}i_1}(x_2...x_{N})
\label{ew}
\ee
with the system volume $V$. This allows to calculate the configurational integral (\ref{q})
exactly by successively integrating $x_1...x_N$
\be
&\tilde Q_N&(\beta)\nonumber\\
&=&\sum\limits_{i_1..i_{N-1}} \int dx_1...dx_N dx_1'...dx_{N-1}' \Psi^*_{i_1..i_{N-1}}(x_1'...x_{N-1}')
\nonumber\\&&\times 
\Psi_{i_1..i_{N-1}}(x_1...x_{N-1})
\nonumber\\ && 
\matrix{
\times & (1+f_{12})(1+f_{13})(1+f_{14})...(1+f_{1N})\cr
\times & (1+f_{21'})(1+f_{23})(1+f_{24})...(1+f_{2N})\cr
\times &... ...\cr
\times & (1+f_{N1'})(1+f_{N2'})(1+f_{N3'})...(1+f_{NN-1'})
}
\nonumber\\&=&
\sum\limits_{i_1..i_{N-1}} \int dx_1'...dx_{N-1}'
\Psi^*_{i_1..i_{N-1}}(x_1'...x_{N-1}')
\nonumber\\ &&
\times V^N {\rm e}^{-N
  \varepsilon_{\{i\}}}\Psi_{i_1..i_{N-1}}(x_1'...x_{N-1}')
\nonumber\\&=&
V^N\sum\limits_{i_1..i_{N-1}} {\rm e}^{-N \varepsilon_{\{i\}}}.
\label{q1}
\ee
This establishes already the complete proof that we can map a
classical $N$-body system on a $N-1$-body quantum system since
(\ref{ew}) is the eigenvalue problem of a $N-1$-body Schroedinger
equation. To see this
we can consider a wavefunction $\xi$ built from the Fouriertransform
of $\tilde \Psi$
\be
\xi(p_1...p_{N-1},t)={\rm e}^{-{i\over \hbar} \sum\limits_{i=1}^{N-1} {p_i^2\over 2
    m_i} t}\tilde \Psi_{i_1...i_{N-1}}(p_1...p_{N-1})
\ee
which obeys the $N-1$-particle Schroedinger equation
\be
&&\left (i \hbar {\partial \over \partial t}-\sum\limits_i^{N-1}
  {p_i^2\over 2 m_i} -\tilde U \right)  \xi=E_{N-1}\xi
\label{s}
\ee
with 
$E_{N-1}\propto V {\rm e}^{-\varepsilon_{\{i\}}}$ and we rewrote the
left hand side of (\ref{ew})  
as quantum potential 
\be
&&<x_1i_1...x_{N-1}i_{N-1}|U|x_1'i_1'...x_{N-1}'i_{N-1}'>
\nonumber\\&=&
{\rm e}^{{\beta}
  [u_{12}(x_{1}'-x_1)+...+u_{1N}(x_{1}'-x_{N-1})]}\nonumber\\&&\times\Omega
\delta(x_1-x_2')...\delta(x_{N-2}-x_{N-1}') \delta_{i_1',i_1}
...\delta_{i_{N-1}',i_{N-1}}.
\label{u}
\ee
The resulting equivalent quantum potential
(\ref{u}) is a $N-1$-body nonlocal potential with respect to the
coordinates but depends on $N$ strength function parameter (e.g.
charges). Therefore we have casted a classical $N$-body problem into a
nonlocal quantum $N-1$ body problem. One could easily give also a
symmetrised or anti-symmetrised form
of the potential using symmetries of the wave function and permuting
coordinates of (\ref{u}) respectively. We do not need it here since we will
restrict to applications neglecting exchange correlations further on.

While the above correspondence holds for any particle number and might
be useful to find solvable models for classical three - body problems,
we will consider in the following many - body systems. First let us
invert the problem and search for an effective classical potential
approximating quantum systems. This should us lead to the known Kelbg-potential
(\ref{r}).
For this purpose we assume a quantum system described in  lowest 
approximation by a Slater determinant or
a complete factorisation of the
many - body wave function  into single wave function
$\Psi_{i_1...i_N}(x_1...x_N)=\phi_{i_1}...\phi_{i_N}$.
We neglect for simplicity exchange correlations in the following.
The corresponding eigenvalue equation for $\phi$ itself one can
obtain from (\ref{ew}) or (\ref{s}) by multiplying with 
$\Psi_{i_2..i_{N-1}}^*(x_2...x_{N-1})$ and integrate over
$x_2...x_{N-1}$. To see the generic structure more clearly we better 
calculate the correlation energy by multiplying (\ref{ew}) or (\ref{s})
by
$\Psi_{i_2..i_{N-1}i_1}^*(x_2...x_N)$ and integrating over
$x_2...x_{N}$. This
provides also the eigenvalue $\epsilon_{\{i\}}$ and leads
easily to approximations for the partition function
(\protect\ref{qn}). To demonstrate this we choose the lowest order
approximation taking identical plane waves for $\phi$. Than the pressure can be
obtained from the partition function $Q_N$ via (\ref{q1}) 
\be
&&P\!=\!T{\partial \over \partial V} \ln Q_N\!=\! T\left ({N\over V}
  \!-\!{N(N\!-\!2)\over V^2} \int \!dr\! \left (\!{\rm e}^{\!-\!\beta u(r)/2}\!-\!1\!\right )
\right )
\nonumber\\&&
\ee
where $V$ is the volume of the system.
We recognise the standard second virial coefficient for small
potentials while for higher order potential the factor $1/2$ appears
in the exponent instead as a pre-factor indicating a different partial
summation of diagrams due to the schema behind (\ref{q1}) and (\ref{s}).

\begin{figure}
\psfig{file=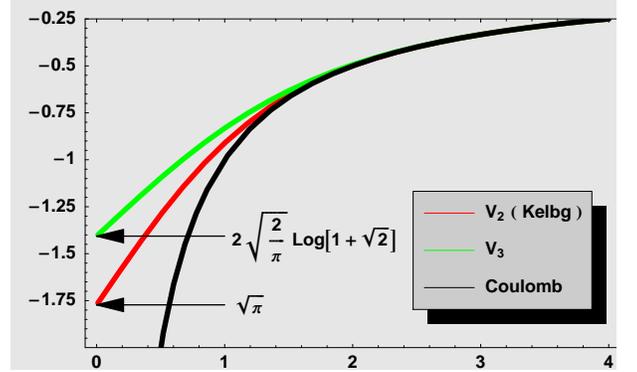,width=10cm,angle=0}
\caption{The comparison of the Kelbg potential (\protect\ref{r}) and
  the third order potential (\protect\ref{15}) versus $r/l$.
}\label{v23a}
\end{figure}

To go beyond the plane wave approximation we
multiply (\ref{ew}) by
$\Psi_{i_2..i_{N-1}i_1}^*(x_2...x_N)$ and the kinetic part of the
statistical operator before integrating over
$x_2...x_{N}$. This means we create an integral over the $N-1$
particle density operator and the potential (\ref{u}) which together represents the
correlation energy. This expression is a successive
convolution between the cluster graphs $f_{ij}$ and the relative 
two - particle correlation function 
$\rho_{i_1i_2}(x_1-x_2)$.
The resulting mean correlation energy 
density reads
\be
&&{U\over V}\!=\!\sum\limits_{\{i\}}\int {dy_1\!...\!dy_{N-1}\over V^{N-1}} 
\rho_{i_1i_2}(y_1)\rho_{i_2i_3}(y_2)...\rho_{i_{N-1}i_1}(y_{N-1})
\nonumber\\&&\times(1\!-\!f_{12}(y_1))(1\!-\!f_{13}(y_1\!+\!y_2))...(1\!-\!f_{1N}(y_1\!+\!...\!+\!y_{N-1}))
\nonumber\\
&&\approx\sum\limits_{\{i\}}\int {dy_1\!...\!dy_{N-1}\over V^{N-1}}
  \rho_{i_1i_2}(y_1)\!...\!\rho_{i_{N-1}i_1}(y_{N-1})
\nonumber\\&&\times
  u_{12}(y_1)u_{13}(y_1\!+\!y_2)\!...\!u_{1N}(y_1\!+\!...\!+\!y_{N-1})+...
\label{u1}
\ee
in dimensionless units where all other cluster expansion terms lead
either to
lower mean field  or disconnected terms. While these terms can be
calculated as well we restrict to the highest order convolutions in
the correlation energy (\ref{u1}) which have now
the structure of mean correlation energy $U/V=\sum\limits_{i_1i_2}\int {dx\over V}
\rho_{i_1i_2}(x) V^{\rm eff}_{12}$ with a classical effective
potential $V^{\rm  eff}_{12}$
\be
 V^{\rm eff}_{2}(r)&\propto& \sum\limits_{3}\int {d x_1\over V} \rho_{12}(x_1) u_{12}(x_1) u_{23}(x_1+r)
\label{v2}
\\
 V^{\rm eff}_{3}(r)&\propto&\sum\limits_{34}\int {d x_1 dx_2\over V^2} \rho_{12}(x_1) u_{12}(x_1)
u_{13}(x_1+x_2) 
\nonumber\\&&\qquad \qquad\times
\rho_{23}(x_2) u_{34}(x_1+x_2+r)
\nonumber\\&&...
\label{v}
\ee
where the two-particle, three-particle etc. approximation can be given.
In equilibrium the nondegenerate correlation function reads $[l^2=\hbar^2/\mu T=\lambda^2/2\pi]$
\be
\rho_{i_1i_2}(x_1\!-\!x_2)\!&=&\!\!\int \!{d p\over (2 \pi \hbar)^3} {\rm e}^{i p r/\hbar} \lambda^3 {\rm
  e}^{-\beta {p^2\over 2\mu}}={\rm e}^{-r^2/l^2}.
\label{rho1}
\ee
Using the Coulomb potential $u\propto 1/r$ we
obtain from the two-particle approximation (\ref{v2}) just the Kelbg 
potential (\ref{r}).
The three - particle
approximation (\ref{v}) can be 
calculated as well and reads $[x=r/l]$
\be
&&V^{\rm eff}_3\sim {1\over x} \left ( {\rm erf}^2\left ({x\over
      \sqrt{2}}\right )\!+\!
  {2^{3/2} x\over \sqrt{\pi}} \int\limits_x^\infty {dz \over z} {\rm
      e}^{-z^2/2} {\rm erf}\left ({z\over \sqrt{2}} \right ) \right ).
\nonumber\\&&
\label{15}
\ee
The comparison of the third order potential with the Kelbg potential
can be seen in figure~\ref{v23a}. The third order potential is
somewhat less bound than the Kelbg potential.
With the schema (\ref{v}) one can easily integrate higher order
approximations as successive convolutions, but with respect to the
small differences between (\ref{r}) and (\ref{15}) in
figure~\ref{v23a} one does not expect much change. 
Also in principle the
degenerate case could be
calculated using Fermi-Dirac distributions in (\ref{rho1}).  But one
should then consider also the neglected exchange correlations during
factorisation of $\Psi$ as well.
Let us summarise that the known effective classical potential
describing a quantum system in binary approximation has been recovered
by identifying  the  effective two-particle 
interaction within the correlation energy.

We want now to proceed to a phenomenological level in that the above
correspondence between
quantum and classical systems motivates to find good approximations
for the dynamics of classical many-body systems by employing quantum-Born
approximations. This can be understood
by the fact that the Kelbg potential deviates appreciably from the
Coulomb one only if the interparticle distance $d$ are smaller than the
thermal wave length $\lambda$. In other words for dense classical
systems under such conditions we can think of it as a dilute quantum
system replacing $\lambda \sim d$.
To check this conjecture let us consider an one-component plasma
system which is characterised by two values. The classical 
coupling is described by the plasma parameter 
$\Gamma={e^2 \over d T}
$
as a ratio of the length where Coulomb energy becomes larger than kinetic
energy, ${e^2 \over T}$, to the interparticle distance or Wigner size
radius $d=({3 \over 4 \pi n})^{1/3}$. Ideal plasmas are found for
$\Gamma<<1$ while around $\Gamma=1$ non-ideal effects become
important. A second parameter which controls the quantum features is
the Bruckner parameter as the ratio of the Wigner size radius to the
Bohr radius $a_B=\hbar^2/m e^2$.
Quantum effects will play a role if $r_s \le 1$. We will consider the
situation that the interaction of such system is switched on at
initial time. Then the correlations are formed by the system which is
seen in an increase of temperature accompanied by the build up of
negative correlation energy. This theoretical experiment has been
investigated numerically by \cite{GZ99} for classical plasmas with 
different plasma parameter $\Gamma$. 

\begin{figure}
\psfig{file=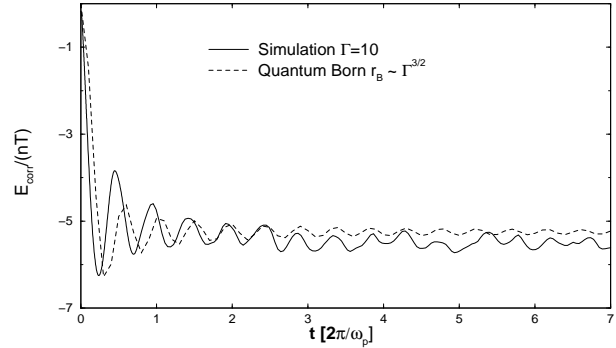,width=8cm,angle=-90}
\caption{The time evolution of a classical one-component plasma after
  sudden switching of interaction \protect\cite{GZ99} 
compared to the quantum Born result
  when the Bruckner parameter is replaced according to
  (\protect\ref{fit}). The long time equilibrium value is remarkably well
reproduced by the quantum - Born result(\protect\ref{eq}).
}\label{hfit}
\end{figure}

In \cite{MSL97a,MK97} we have calculated the formation of such
correlations by using quantum kinetic equations in Born approximation.
The time dependence of kinetic energy was found at short times to be
\begin{eqnarray}
E_{\rm corr}&=&-\sum_{ab}\int\frac{dkdpdq}{(2\pi\hbar)^9}V^2_{\rm D}
\frac{1-\cos\left\{{1\over\hbar}t\Delta_E\right\}}{\Delta_E}
\nonumber \\
&&\times f'_a f'_b(1-f_a)(1-f_b)
\label{energ1}
\end{eqnarray}
where $f$ are the initial distributions and $\Delta_E={k^2\over 2m_a}+{p^2\over 2m_b}-{(k-q)^2\over 2m_a}-
{(p+q)^2\over 2m_b}$. The statical screened Coulomb interaction 
is $V_D(q)=4 \pi e^2 \hbar^2/(q^2+\hbar^2 \kappa^2)$
with the inverse screening length
expressed by density $n$ and temperature $T$ as 
$\kappa^2=4 \pi e^2 n /T$ or  
$\kappa^2=6 \pi e^2 n/\epsilon_f$ 
for the high or low
temperature limit. 
For both cases dynamical as well as statical
screening it was possible to integrate analytically the time dependent
correlation energy (\ref{energ1}).
This has allowed to describe the time dependence of simulations in the
weak coupling limit $\Gamma<1$
appropriately \cite{MSL97a}. For stronger coupling $\Gamma \ge 1$ the Born
approximation fails since the exact correlation energy of
simulation is lower
than the first order (Born) result $\kappa e^2/2
T=\sqrt{3/2}\Gamma^{3/2}$. 
Moreover there appear typical oscillations as seen in figure~\ref{hfit}.

Now we will employ the ideas developed above and will use the quantum
Born approximations in the strongly degenerated case to describe the
classical strongly correlated system. For strongly degenerated plasmas
the time dependence of correlation energy was possible to integrate as
well with the result \cite{MK97} expressed here in terms of plasma
parameter $\Gamma$ and quantum Bruckner parameter $r_s$ as
\be
{E_{\rm corr}^T(t)-E_{\rm corr}^0(t)\over n T}&=&
{1\over (36 \pi^4)^{1/6}} {r_s^3\over
  \Gamma}  \left ( {\sin{y \tau}\over y \tau}-1\right )
\nonumber\\&&\times
\left ( {1\over b_l} {\rm arctan}({1 \over b_l})+{1\over b_l^2+b_l^4} \right
    )
\label{eq}
\ee
with $b_l=\hbar \kappa/2 p_f=\sqrt{\Gamma}/(48 \pi^2)^{1/6}$, $y \tau=
4 \epsilon_f t/\hbar=(2)^{4/3} \pi^{5/3} 3^{5/6}  \tau/\sqrt{r_s}$
where the time is scaled in plasma periods $\tau =2 \pi t/\omega_p$.
Now we try to fit this quantum result to the simulation using the
Bruckner parameter as free parameter. For the available simulations
between $1 \le \Gamma \le 10$ we obtain a best fit
\be
r_s^{\rm fit}= c \sqrt{3\over 8} \Gamma^{3/2} \qquad c\approx 0.5.
\label{fit}
\ee
The quality of this fit is illustrated in figure~\ref{hfit} which is
throughout the range $1 \le \Gamma \le 10$. This is quite astonishing
since not only the correct classical correlation energy \cite{I94} is described
but also the correct time dependence i.e. dynamics.

Let us try to understand what this phenomenological finding
means. Using the thermal De Broglie wave length
$\lambda^2=\hbar^2/4 m T$ we can rewrite (\ref{fit}) as
\be
{\lambda^2}\approx {d \over \kappa}={d^2\over (3 \Gamma)^{1/2}}.
\ee
In the considered range of $\Gamma=1...10$ we have $(3
\Gamma)^{1/4}=1...2$ and the thermal wave length $\lambda$ is found to be nearly equal 
the interparticle distance $d$ as a best fit of quantum Born
calculation to dense classical systems. This is exactly the distance
where the Kelbg potential (\ref{v2}) or (\ref{r}) starts to deviate from the
Coulomb potential. In other words we confirm the conjecture that the
dense classical system can be described by dilute quantum systems if
in the latter systems  the thermal wave length is replaced by the interparticle
distance.
This condition (\ref{fit}) can also be rewritten into the result (\ref{1}) of
literature using the degenerated screening length. 

We summarise that in
equilibrium we have shown that there exist an exact relation between
a $N$-body classical system and a $N-1$-body quantum system. This has
allowed to recover the quantum Kelbg potential easily.  
As practical consequence we suggest to
describe the dynamics of dense interacting classical many body systems
by the simpler 
perturbative quantum calculation in degenerate limit replacing
properly $\hbar$ by typical classical parameters of the system.

I would like to thank S. G. Chung  for numerous
discussions and valuable hints.


\begin{thebibliography}{10}

\bibitem{BMM00}
T.~S. Bir{\'o}, S.~G. Matinyan, and B. M{\"u}ller,   (2000), hep-th/0010134.

\bibitem{Ch99}
S.~G. Chung, Phys. Rev. B {\bf 60},  11761  (1999).

\bibitem{BH84}
D. Bohm and B.~J. Hiley, Foundations of Physics {\bf 14},  255  (1984).

\bibitem{MM00}
V.~I. Man{'}ko and R.~V. Mendes, Physica D {\bf 45},  330  (2000).

\bibitem{K64}
G. Kelbg, Ann. Physik {\bf 13},  354  (1964).

\bibitem{K641}
G. Kelbg, Ann. Physik {\bf 14},  394  (1964).

\bibitem{EHK67}
W. Ebeling, H. Hoffmann, and G. Kelbg, Beitr{\"a}ge aus der Plasmaphysik {\bf
  7},  233  (1967).

\bibitem{KK68}
D. Kremp and W.~D. Kraeft, Ann. Physik {\bf 20},  340  (1968).

\bibitem{kker86}
W.~D. Kraeft, D. Kremp, W. Ebeling, and G. R{\"o}pke, {\em Quantum Statistics
  of Charged Particle Systems} (Akademie Verlag, Berlin, 1986).

\bibitem{KK681}
W.~D. Kraeft and D. Kremp, Zeit. f. Physik {\bf 208},  475  (1968).

\bibitem{OVE00}
J. Ortner, I. Valuev, and W. Ebeling, Contrib. Plasma Phys.  .

\bibitem{GZ99}
G. Zwicknagel, Contrib. Plasma Phys. {\bf 39},  155  (1999).

\bibitem{MSL97a}
K. Morawetz, V. {\v S}pi{\v c}ka, and P. Lipavsk{\'y}, Phys. Lett. A {\bf 246},
   311  (1998).

\bibitem{MK97}
K. Morawetz and H. K{\"o}hler, Eur. Phys. J. A {\bf 4},  291  (1999).

\bibitem{I94}
S. Ichimaru, {\em Statistical Plasma Physics} (Addison-Wesley Publishing
  company,, Massachusetts, 1994), p.\ 57.

\end{thebibliography}

\end{document}